\documentclass[aip,jmp,amsmath,amssymb]{revtex4-1}

\usepackage{graphicx}
\usepackage{dcolumn}
\usepackage{bm}
\def\0#1{{\mathrm{#1}}}
\def\1#1{{\mathbb{#1}}}
\def\2#1{{\mathbf{#1}}}
\def\3#1{{\mathcal {#1}}}
\def\4#1{{{\mathsf{#1}}}}
\def\5#1{{{\widetilde{#1}}}}  
\def\6#1{\overline{#1}}
\def\7#1{\breve{#1}}
\def\8#1{{\widehat{#1}}}
\def\9{\tiny}

\def\<{{\left<{}\right.}}
\def\>{{\left.{}\right>}}
\def\O{\mbox{ \o\kern 4pt}}

\def\x{\times}

\def\W{{\bigwedge}}
\def\w{\mathop{\wedge}\nolimits}

\def\BAR{\begin{array}}
\def\EAR{\end{array}}
\def\BEQ{\begin{equation}}
\def\EEQ{\end{equation}}
\def\BEN{\begin{enumerate}}
\def\EEN{\end{enumerate}}
\def\BIT{\begin{itemize}}
\def\EIT{\end{itemize}}
\def\BEA{\begin{eqnarray}}
\def\EEA{\end{eqnarray}}
\def\BED{\begin{description}}
\def\END{\end{description}}
\def\BET{\begin{table}}  
\def\ENT{\end{table}}
\def\apo{\mbox{'}}
\def\cliff{\mathop{{\;\rule[-2pt]{.6pt}{9pt}
{\kern-1.3pt}
\rule[7pt]{7pt}{1pt}
{\kern-1.3pt}
\rule[-2pt]{.6pt}{9pt}}\,}\nolimits}

\def\dim{\mathop{{\mathrm{dim}}}\nolimits}
\def\dual{\mathop{{\mathrm {dual}}}\nolimits}
\def\dup{\mathop{{\mathrm {dup}}}\nolimits} 

\def\hexp{\mathop{{\mathrm{hexp}}}\nolimits}
\def\lbl{\label}

\def\oo #1{{\stackrel{\circ}{#1}}}
\def\op{{\mathop{\mathrm{op\,}}\nolimits}}

\def\go{\mathrel{\stackrel{>}{\scriptstyle \sim}}}

\def\oar{{\rightharpoonup}}

\def\rao{{\leftharpoonup}}
\def\Qi{{\oo{\imath}\,}}

\def\so{\mathop{{\mathrm {so}}}\nolimits}

\def\SO{\mathop{{\mathrm {SO}}}\nolimits} 

\def\spin{\mathop{{\mathrm {spin}}}\nolimits}

\def\su{\mathop{{\mathrm {su}}}\nolimits} 

\def\SU{\mathop{{\mathrm {SU}}}\nolimits} 

\def\to{\rightarrow}
\def\Z{{\vrule width2pt height0pt depth0pt}}
\def\x{\times }

\begin{document}

\title[Isospin, color, generation]{Modular architecture of isospin, color, and generation}
\author{David Ritz Finkelstein}
\email{finkelstein@gatech.edu}
\affiliation{Georgia Institute of Technology}

\date{\today}

\begin{abstract}
Starting from the vacuum, iterated Grassmann-algebra formation consecutively introduces
vector spaces and groups with the structure first of charge, then of isospin, then of color.  \end{abstract}
\maketitle
\section{\lbl{S:CORE} Quantum set theory} 

The cellular model 
of the internal variables of the Standard Model constructed here 
began as a finite quantum relativistic
version of the von Neumann cellular automaton
\cite{FINKELSTEIN1969,FINKELSTEIN1969a}.
The usual language for the description of finite systems
is finite set theory.
But set theory incorporates Boolean algebra,
and quantum theory revises Boolean algebra at the quantum level,
so that quantum theory must also revise set theory at that level.
A quantum set theory is reviewed here and applied to the internal variables of the Standard Model.

Quantum sets turn out to be spins
of associated orthogonal groups.
Therefore
this work is also in the line
of earlier proposals of Feynman  \cite{FEYNMAN1941}
and Penrose \cite{PENROSE1971}, 
composing the world out of spins.
It adds the key element of
{\em modular architecture}
\cite{SIMON1962}, from which a world of quantum bits or spins arises naturally.
I recapitulate quantum set theory (Section \ref{S:QSETTHEORY}),
 and apply it to isospin and color (Section \ref{S:COLOR}), generation (Section \ref{S:GENERATION}), and orbital variables
(Section \ref{S:FERMIONRANK}).

\section{\lbl{S:QSETTHEORY} Quantum set theory}

A {\em quantum set} is a Fermi-Dirac (or {\em odd}) combination of identical quantum systems. More fully put:

Let $\3V$ be the input
vector space of a quantum system \cite{FOCK1931}.
Let $\iota$ be the unitizer:
\BEQ
\iota v :=\{v\}\quad \mbox{for all }v\in \3V,
\EEQ
modulo the identification 
\BEQ
\iota(av+v')\equiv a\iota v + \iota v'
\EEQ
that makes the mapping $\iota:\3V\to \iota\apo \3V$ linear.
Let $\iota\apo \3V$ designate the vector space of unit sets of
elements of $\3V$.
Let $\W\iota\apo \3V$ be the Grassmann algebra over
$\iota\apo \3V$.
Let $\3S$ be the Grassmann algebra over its own unitization 
that is minimal in that regard:
\BEQ
\3S=\W\iota\apo \3S = \bigcup_{r=1}^{\infty}\3S^r,\quad \3S^r:=\left[\W\iota\apo\right]^r\1C.
\EEQ
The classical correspondent of $\3S$ is  the set 
$\3S_{\0c}$
of all sets
of finite rank and empty foundation.

Then the {\em quantum set of rank} $r$  
is the hypothetical quantum system with input vector space $\3S^r\subset \3S$.
Ranks nest: $\3S^0 \subset \3S^1\subset \3S^2\subset \dots $\/.

Suppose that the history from input to output of the quantum system under study  is a quantum set of some rank $R$ to be determined.
Provisionally, $R\approx 6$ suffices for quantum field theories (Section \ref{S:FERMIONRANK}).

This  gives algebraic expression, for example, to statements that
a fermion field {\em includes} ($\supset$) elementary fermions,
and that an elementary fermion {\em contains} ($\ni$) a spin, an isospin, and a color.

The modules of this modular architecture are the unit quantum sets of rank $\le R$.

The input space of the generic quantum set is infinite-dimensional, 
but  modular quantum theory currently stops at rank 6, 
whose dimensionality is only 
\BEQ
d= 2^{(2^{64\0K})}.
\EEQ

The non-semisimple algebras and infinite spectra of present physics 
arise in quantum set theory only as singular limits.

Table \ref{T:QSETS} below samples a basis of quantum sets,
the {\em standard basis},
and assigns serial numbers to its basic sets
in the order of generation.  
\begin{table}[h]
\begin{center}
\begin{tabular}{|r|cccccccccccc|cr|}
\hline
$\vdots$ &&&& && &$\vdots$&&&& &&
&\cr
\hline
6 & \vrule height 18pt depth 0pt width 0pt
$\6{\6{\6{\6{\6{\6{\Z}}}}}}$&
$\6{\6{\6{\6{\6{\6{\Z}}}}}}\,{\6{\Z}}$&
$\6{\6{\6{\6{\6{\6{\Z}}}}}}\,{\6{\6{\Z}}}$&
$\6{\6{\6{\6{\6{\6{\Z}}}}}}\,\6{\6{\Z}}\,\6{\Z} $&
$\6{\6{\6{\6{\6{\6{\Z}}}}}}\,\6{\6{\6{\Z}}}$&
$\6{\6{\6{\6{\6{\6{\Z}}}}}}\,\6{\6{\6{\Z}}}\,\6{\Z}$&
$\6{\6{\6{\6{\6{\6{\Z}}}}}}\,\6{\6{\6{\Z}}}\,\6{\6{\Z}}$&
$\6{\6{\6{\6{\6{\6{\Z}}}}}}\,\6{\6{\6{\Z}}}\,\6{\6{\Z}}\,\6{\Z}$&
$\6{\6{\6{\6{\6{\6{\Z}}}}}}\,\6{\6{\6{\Z}}\,\6{\Z}}$&
${}{\cdots}$&
$\6{\6{\6{\6{\6{\6{\Z}}}}}\,\6{\Z}}$&
$\6{\6{\6{\6{\6{\6{\Z}}}}}\,\6{{\Z}}}\,\6{\Z}$&
${}{\cdots}$
&\cr
& \vrule height 14pt depth 0pt width 0pt
$2^{(2^{16})}\!\!\!$&\dots&&&&&&&&\dots&&&\dots
&\cr
\hline
5 & \vrule height 14pt depth 0pt width 0pt
$\6{\6{\6{\6{\6{\Z}}}}}$&
$\6{\6{\6{\6{\6{\Z}}}}}\,\6{\Z}$&
$\6{\6{\6{\6{\6{\Z}}}}}\,\6{\6{\Z}}$&
$\6{\6{\6{\6{\6{\Z}}}}}\,\6{\6{\Z}}\,\6{\Z} $&
$\6{\6{\6{\6{\6{\Z}}}}}\,\6{\6{\6{\Z}}}$&
$\6{\6{\6{\6{\6{\Z}}}}}\,\6{\6{\6{\Z}}}\,\6{\Z}$&
$\6{\6{\6{\6{\6{\Z}}}}}\,\6{\6{\6{\Z}}}\,\6{\6{\Z}}$&
$\6{\6{\6{\6{\6{\Z}}}}}\,\6{\6{\6{\Z}}}\,\6{\6{\Z}}\,\6{\Z}$&
$\6{\6{\6{\6{\6{\Z}}}}}\,\6{\6{\6{\Z}}\,\6{\Z}}$&
${}{\cdots}$&
$\6{\6{\6{\6{\6{\Z}}}}\,\6{\Z}}$&
${}{\cdots}$&
&\cr
& \vrule height 14pt depth 0pt width 0pt
$2^{16}$&\dots&&&&&&&&\dots&&\dots&
&\cr
\hline
4 & \vrule height 14pt depth 0pt width 0pt
$
\6{\6{\6{\6{\Z}}}}$&$
\6{\6{\6{\6{\Z}}}}\,\6{\Z}$&$
\6{\6{\6{\6{\Z}}}}\,\6{\6{\Z}}$&$
\6{\6{\6{\6{\Z}}}}\,\6{\6{\Z}}\,\6{\Z} $&$
\6{\6{\6{\6{\Z}}}}\,\6{\6{\6{\Z}}}$&$
{{\6{\6{\6{\6{\Z}}}}\,{\6{\6{\6{\Z}}}\,\6{\Z}}\,}}$&$
{{\6{\6{\6{\6{\Z}}}}\,\6{\6{\6{\Z}}}\,\6{\6{\Z}} }}$&$
{{\6{\6{\6{\6{\Z}}}}\,\6{\6{\6{\Z}}}}\,\6{\6{\Z}}\,\6{\Z}}$&$
\6{\6{\6{\6{\Z}}}}\,\6{\6{\6{\Z}}\,\6{\Z}}$&$
{}{\cdots}$&$
{{\6{\6{\6{\6{\Z}}}\,\6{\Z}}}}$&
$\cdots$&
15 possible&\cr
& \vrule height 14pt depth 0pt width 0pt
16&17&18&19&20&21&22&23&24&\dots&32&$\dots$&
generations&\cr
\hline
3 & \vrule height 14pt depth 0pt width 0pt
$
{\6{\6{\6{\Z}}}}$&$
{{\6{\6{\6{\Z}}}\,\6{\Z}}}$&$
{{\6{\6{\6{\Z}}}\,\6{\6{\Z}}}}$&$
{{\6{\6{\6{\Z}}}\,\6{\6{\Z}}\,\6{\Z}}}$&$
{{\6{\6{\6{\Z}}\,\6{\Z}}}}$&$
{{\6{\6{\6{\Z}}\,\6{\Z}}\, \6{\Z} }}$&$
{\6{\6{\6{\Z}}\,\6{\Z}} \, \6{\6{\Z}}\,}$&$
{\6{\6{\6{\Z}}\,\6{\Z}}\,\6{\6{\Z}}\,\6{\Z}}$&$
{\6{\6{\6{\Z}}\,\6{\Z}}\,\6{\6{\6{\Z}}}}$&$
{\,\6{\6{\6{\Z}}\,\6{\Z}}\,\6{\6{\6{\Z}}}\,\6{\Z}}$&$
{\6{\6{\6{\Z}}\,\6{\Z}}\,\6{\6{\6{\Z}}}\,\6{\6{\Z}}}$&$
{{\6{\6{\6{\Z}}\,\6{\Z}}\,\6{\6{\6{\Z}}}\,\6{\6{\Z}}\,\6{\Z}}}$&
$\{Q_L, \6{Q_R}\}$
&\cr
& \vrule height 14pt depth 0pt width 0pt
4&5&6&7&8&9&10&11&12&13&14&15&
&\cr
\hline
2 & \vrule height 14pt depth 0pt width 0pt ${\6{\6{\Z\Z}}}$&$
{{\6{\6{\Z\Z}}\,\6{\Z\Z}}}$& &&&&&&&&&&
$\{e_L,\nu_L\}$
&\cr
& \vrule height 14pt depth 0pt width 0pt 2&3& &&&&&&&&&&
&\cr
\hline
1 & \vrule height 14pt depth 0pt width 0pt${\6{1}}$&
&&&&&&&&&&&
$\6{e_R}$
&\cr
 & \vrule height 14pt depth 0pt width 0pt${1}$&
&&&&&&&&&&&
&\cr
\hline
0 & \vrule height 14pt depth 0pt width 0pt  1&
&&&&&&&&&&&
$\6{\nu_R}$
&\cr
 & \vrule height 14pt depth 0pt width 0pt  0&
&&&&&&&&&&&
&\cr
\hline
\hline
$r$&$s_n$&&&&&&&&&&&&
kinds
&\cr
&$n$&&&&&&&&&&&&

&\cr
\hline
\end{tabular}
\vskip20pt
\end{center}
\caption{\lbl{T:QSETS}{\bf Basic sets and fermion kinds.}
$r=$ proper rank,
$s_n=$ $n$th basic set vector.}
\end{table}

Beneath each set $s_n$ is its {\em serial number} $n$.

Table \ref{T:QSETS} can also serve as a table of monads (unit sets, 
first-grade Grassmann elements) if a bar is imagined over every entry.
This is useful since the fundamental fermions and their modules are monads.

Elements of the standard basis of rank $r$ that are 
not elements of rank $r-1$ are said to have proper rank $r$
and belong to  $\3S^{[r]}$.
Unlike ranks, proper ranks are disjoint, 
except for the common element 0.

I cast Cartan's construction of spinors into the present terms:

For any complex vector space $\3V$, the dual space $\dual \3V:=\5{\3V}$
is the space of linear mappings $\3V\to \1R$.
The duplex space is
\BEQ\lbl{E:DUP}
\dup \3V:= \dual \3V \oplus \3V.
\EEQ
Define the {\em duplex form} $\|w\|_{\dup}$ on $\3W=\dup \3V$
by the condition
\BEQ
\forall v\in \3V,\,u\in\5{\3V}:\quad  
\|u+v\|_{\dup}:=u\circ v,
\EEQ
the value of $u$ on $v$.
The duplex form is neutral. 
It is invariant under a group $\SO(2n\1C)$ of complex matrices,
where $n=\dim \3V$,
but this group is reduced by the condition (\ref{E:DUP}).

Define the {\em spinors} of $\so(2n)$ as the elements of 
$\W \3V$. 
They support the spin representation of $\so(2n\1C)$.

The vectors of the Grassmann algebra $\3S^r$ are thus spinors of the real Lie algebra
$\so(\dup \3S^{r-1})$.
There is thus a tautologous connection between statistics and spin, but the spin is not restricted to the Lorentz group yet.

\section{\lbl{S:COLOR} Isospin and color}

The 16 monads of rank 4 in
Table \ref{T:QSETS} are clearly partitioned into four proper ranks 1, 2, 3, and 4, 
with respective multiplicities 1, 1, 2, and 12.
The contents of these monads are all the entries of proper ranks 0, 1, 2, and 3 respectively.

The 16 fundamental left-handed fermions of the Standard Model
have a similar pattern: the left-handed antineutrino
is one isospin singlet,
the left-handed positron is a second,
the left-handed neutrino and the left-handed electron 
form an isospin doublet,
and 12 left-handed quarks repeat this pattern
in three colors, $12 = 3\x (1+1+2)$.

This suggests a structural correspondence between fermion kinds
and monads of Table \ref{T:QSETS}:
\BEN\lbl{I:CHROMON}
\item [C1.] The two lepton singlets correspond
to the two monads of proper ranks 2 and 1.
\item[C2.] The lepton isospin doublet corresponds to the two monads of proper rank 3.
\item[C3.] The 12 quarks correspond to the 12 monads of proper rank 4.
\EEN
These fermions are entered next to the contents of their monad
in Table \ref{T:QSETS}, not the monads themselves.
The 24-dimensional Standard Model internal group,
\BEQ
\3G_{\0{SM}}:=\SU(3)\x \SU(2)\x \0U(1)
\EEQ
then acts on $\3S^3$ as follows.

Proper rank 2, being two-dimensional, defines an $\SU(2)$ group
 that
blends the two basis vectors of proper rank 2;
that is,
maps each into a superposition of both.
Since according to the assignment
C1, proper rank 2 represents a lepton isospin doublet,
this proper-rank-2 $\SU(2)$ must be identified with isospin $\SU(2)$.

If this assignment is correct,
the 12 monadic dimensions of quark proper rank $\3S^{[4]}$ 
should then resolve into 3 copies of lepton space $\2\1C=1+1+2$,
 one for each color, 
coherently mixed by a color $\SU(3)$ group.

Proper rank $\3S^{[4]}$ consists of polynomials in the unitizations of the vectors of rank 3.  

Assume that $\iota:\3S^{[3]}\to \3S^{[4]}, \3S^{[4]}\to \3S^{[5]}$ 
commutes with $\SU(2):\3S^{[3]}\to \3S^{[3]},
\3S^{[4]}\to \3S^{[4]}$.

This fixes the action of $\SU(2)$ on $\3S^{[4]}$.

Inspection of Table \ref{T:QSETS} shows that the $12=3\x 4$
monads of proper rank 4 are constructed by inserting any one of the three
elements $s_{4},s_{8},s_{12}$ of proper rank 3 into any one of the four lepton monads
of rank 3.
The three vectors 
\BEQ
c_1:=s_{4},\quad c_2:=s_{8},\quad c_3=s_{12}
\EEQ 
are distinguished
algebraically as the non-constant monomials
in the unit sets of proper rank 3 only.
The constant monomial is 1; 
its insertion into the lepton merely reproduces the lepton.

Therefore it is permissible to identify $s_{4},s_{8},s_{12}$
with three independent input vectors for 
a {\em chromon}---``piece of color",
a quantum module that when inserted into a lepton module
converts it to a quark module.  
It is not a particle, since it lacks orbital variables. 
It is two ranks below the particle rank in
the hierarchy of rank. 
It has no operators but those of the color algebra
$\1C(3)$.

Infer that a quark is an elementary fermion 
that contains a quark module, 
and that a quark module is a lepton module 
with a chromon inserted.
A chromon is an element of a quark, 
not a part of it. 
A quark is still an elementary particle, with no parts.

Now it is meaningful to ask whether the strong interactions proceed by chromon exchange, and whether
leptons are inert to strong interactions because they contain
no chromons to exchange.

A gluon cannot be a unit quantum set, which has odd statistics.
It can be a pair, or a quadruple, or $\dots$.
If a gluon is a quark pair
\cite{SALLER1974}, two gluons might bind by chromon exchange.
The interaction between a gluon and a quark might likewise be
mediated by chromon exchange. 
Such possibilities need further study.

The space $\3S^{[3]}$ indeed splits into three isomorphs of $3S^2$.
For these to be the three quark color spaces,
they should be superposed by a color $\SU(3)$ group that commutes with the proposed isospin $\SU(2)$. 

In fact the color Lie algebra $\su(3)$ is that generated by the eight antihermitian  traceless sesquilinear combinations 
of chromon input operators $c_i$---creation operators---and 
their dual output operators $\6c^i$---annihilation operators:
\BEQ 
c_1\6c^1-\6c^1c_1,\quad c_1\6c^2-\6c^2 c_1, \quad
i(c_1\6c^2+\6c^2 c_1), \quad  \dots
\EEQ

\section{\lbl{S:GENERATION} Generation}

The modular structure iterates.
Just as the monads of proper rank 4 consist of $2^2-1=3$ isomorphs of 
the $2^2$ monads of proper rank 3
(the first-generation quark kinds),
the monads of proper rank 5 consist of $2^4-1=15$ isomorphs 
of the $2^4=16$ monads of rank 4.
They are generated by inserting, into each of the 
16 monads of rank 4 in turn, 
each of the $2^{4}-1=15$ non-constant monomials 
in the four monads of proper rank 4.

Models of generation being scarce, 
it is tempting to infer, at least provisionally and tentatively,
that these 15 copies include the two observed higher generations
of the first-generation fermion kinds:

\BEN\lbl{I:CHROMON4}
\item[C4] The three generations of fundamental fermions
correspond to three of the 16 isomorphs of the rank-4 monads in the rank-5 monads.
\EEN

Since only three generations are found experimentally,
one must then infer that the other 13 kinds are too massive to produce
or too unstable to be detected.
Perhaps the three monomials of rank 4 that contribute their 
genes to the three generations are the first three, 
\BEQ
s_{16}=\iota^4 1, \quad  s_{32}= \iota(\iota^3 1 \w \iota 1),
\quad s_{48}=s_{32}s_{16}
\EEQ
merely on the grounds that they are the simplest.

\section{\lbl{S:FERMIONRANK} The rank of the fermion}

$\3S^r$ has complex dimension designated by $\hexp r$:
\BEQ
\BAR{rcccccccc}
  \dim\3S^r=&1 &2& 4& 16& 2^{16}& 2^{(2^{16})}&2^{(2^{(2^{16})})}&\dots\cr
\mbox{for\hskip13pt   $r=$}&0& 1& 2& 3& 4& 5& 6& \dots
\EAR
\EEQ
The corresponding table for the number of monads of rank $r$
is
\BEQ
\BAR{rcccccccc}
  \dim\3S_1^r=&1 &2& 4& 16& 2^{16}& 2^{(2^{16})}&2^{(2^{(2^{16})})}&\dots\cr
\mbox{for\hskip13pt   $r=$}& 1& 2& 3& 4& 5& 6& 7 & \dots
\EAR
\EEQ

If $L$ in time units is the Planck time $T_{\0P}=5\x 10^{-44}$ s and $T_{\0U}$ is the present age of the universe, $T_{\0U}\approx 4\x 10^{17}$ s, 
then the observed age in Planck units is $N \sim 10^{60}$. 
The number of Planck cells in a history is roughly $10^{240}$.

Thus $\3S^5$ is much too small for field space;
while $\3S^6$ is much larger than present physics needs for field space. 
Despite our abysmal ignorance of the microcosmos,
it is reasonable to conclude that
fermion field space has rank $\go 6$ and arbitrary grade.
If rank 5 is used inefficiently for the generations,
the particle rank may be raised to at least 7.

\section{\lbl{S:CELLULAR}  Assembling cells}
The quantification process defined in this section is used to assemble
space-time from cells in the next section.

Every element of $\3S^{r+1}$ is 
an essentially unique Grassmann polynomial
in the unitizations of the basis elements of $\3S^{4}$.
Therefore every infinitesimal operator $S\in \op \3S^{r}$
has a well-defined representation   
in $\op \3S^{ r+1}$,
called its {\em cumulation} and
written $\sum^{r+1}_r \,S$, defining 
a homomorphism $\sum^{r+1}_4: \op \3S^{ r}\to\op S^{ r+1}$.

Thus every operator in $\op \3S^3$ has a cumulation in
$\op S^4$, and so forth to the particle rank $\op S^R$,
defining a sequence of Clifford homomorphisms
\BEQ
\3C^0\stackrel{\sum^1_0}{\to} \3C^1\stackrel{\sum^2_1}{\to} \dots \stackrel{\sum^{R-1}_{R-2}}{\to}
\3C^{R-1}\stackrel{\sum^R_{R-1}}{\to} \3C^R.
\EEQ
Write their product from rank $r$ to $r'$ as
\BEQ\lbl{E:SUM}
\sum^{r'}_r= 
\sum^{r'}_{r'-1}\;\cdots\; \sum^{r+2}_{r+1}\;\sum^{r+1}_r\/:
\op \3S^r\to \op \3S^{r'}.
\EEQ
Write $\Sigma$ for the Lie algebra homomorphism
induced by the Clifford homomorphism $\sum$.
If $S$ is a spin component in $\op \3S^r$, 
$\Sigma^{r'}_r S$ is the 
sum of corresponding components of all the spins 
of the cells making up the set of rank $r'$,
multiplied as usual by unwritten identity operators.

\section{\lbl{S:ORBITAL} Orbital module}

Canonical quantization enormously reduced the degree of singularity of the algebra of observables of classical mechanics
and field theory, as measured by the radical of 
the Lie algebra.
Yang's space-time quantization completely eliminated the remaining radical.
In Yang's quantized space-time $\3Y$, which is actually a quantized phase space, the Poincar\'e group is reformed to  $\spin(5,1)$ 
or $\spin(3,3)$ \cite{YANG1947}.
Other quantum commutation relations have been proposed 
for orbital variables, but Yang's Lie algebra is the nearest simple one to the standard  Lie algebra $\4a(x,p,L,i)$ of canonical relativistic quantum mechanics \cite{SEGAL1951}.

Yang represents this simple Lie algebra, however, by differential operators on a Hilbert space of functions
on an underlying 6-dimensional continuum $\{\xi^{\alpha}\}$.
The Lie algebra of this Hilbert space is still singular.

To regularize the representation too,
use spin operators
instead of differential operators.
The $\spin(3,3)$  spinors have 8 components,
4 of left-handed chirality and 4 of right-handed.
Identify them with the spinors in a suitable 8-dimensional subspace of the 16-spinor space $\3S^3$;
and
single out a 10-dimensional subalgebra $\4a_{\0Y}\subset \4a(\3S^3)$ of the 16-dimensional algebra
$\4a(\3S^3) \cong \spin(4,4)$  of $\3S^3\sim 16\1C$,
leaving the meaning  of the other six dimensions
for later.


$\3S^3\sim 16\1C$ is the quantum space of the prototypical unit cell of the world cellular automaton.
Write the quantum unit of time 
as $t_{\0Q}$. 
Dynamical considerations suggest that $t_{\0Q}\gg t_{\0P}$,
the Planck time.
Use $c\hbar t_{\0Q}$ units unless otherwise noted.

Then the spin angular momentum in $\op \3S^3$ is
\BEQ
S_{n'n}= \frac{ [\gamma_{n'},\gamma_n]}2,\quad
\gamma_n \in \op \3S^3\/, n,n'=1,\dots, 8\/.
\EEQ
and obeys the $\so(N_+,N_-)$ Lie-algebra relations
\BEQ
[S_{n'''n''},S_{n'n}] =
-S_{n'''n'}g_{n''n}
+S_{n'''n}g_{n''n'}
+S_{n''n'}g_{n'''n}
-S_{n''n}g_{n'''n'}
\EEQ
Suppose
that the fundamental fermion is a unit set of rank 6, as Section \ref{S:FERMIONRANK} indicates,
 founded on a cell of rank 3:
\BEQ
S_{n'n}:= \2{\sum}^6_3 \;\frac{ [\gamma_{n'},\gamma_n]}2\in \op \3S^6\/.
\EEQ
 
Yang \cite{YANG1947} reformed the commutator Lie algebra of space, time, energy, momentum, angular momentum, boosts, and the complex numbers into
one simple {\em orbital algebra} with symmetry group $\spin(5,1)$
or $\spin(3,3)$.
Call the chosen symmetry group of these two the {\em Yang group}.
For brevity take the Yang group to be $\spin(3,3)$ for the present.
The peculiar three timelike dimensions reduce to one
in the organization of the complex plane.

Following Feynman and Yang,
quantize Yang orbital space further by taking
its orbital variables $S_{n'n}\in \op \3S^6$ 
to be components of the 
cumulative $\spin(4,4)$ angular momentum $S_{n'n}$ of the
 fundamental fermion,
 rather than differential operators.
 
Write $a \rao \oo a$ and $\oo a\oar a$ when the 
standard construct $a$ is a singular limit of the regular construct
$\oo a$ of quantum set algebra. 
The limiting process in general including a self-organization that restricts quantum vectors to some sector of the quantum vector space.

Let  the Clifford vector with components $\gamma^n\in \op \3S^6$ 
obey the Clifford Clause
\BEQ\lbl{E:CLIFF}
\{\gamma^{n'},\gamma^n\}=2g^{n'n},
\quad \mbox{where}\quad n,n'=1,2,\dots 6\/.
\EEQ
Set as usual
\BEQ
\gamma^{n'n}:=\frac 12 [\gamma^{n'},\gamma^n]\in\3T^3\/.
\EEQ
Then in an adapted spinor frame,
 the {\em regularized Yang space}
$\oo{\3Y}$ has the following basic orbital operators on $\3S^6$,
expressed as spin operators 
with $\hbar$, $c$, and the fundamental time $\tau$ as units:
\BEQ \lbl{E:X}
\BAR{lrrl}
S^{n'n} &:=&\frac 12 &\!\!\!\sum_{4}^{6}\;\gamma^{n'n},\cr
\oo p_{m}  &:=& \frac 1{\4N}&\!\!\! S_{5m},\cr
\oo x{}^{m}&:=&&\!\!\!S^{6m},\cr
\Qi         &:=&\frac 2{\4N}&\!\!\!S^{65},\cr
\oo L_{m'm}&:=&&\!\! S_{m'm};
\EAR
\EEQ
here $\sum$ is the iterated quantifier defined in (\ref{E:SUM});
$S_{n'n}$ is the quantified angular momentum tensor of
the representation of the cell algebra $\spin(4,4)$ supported by $\3S^6$;
and $\4N/2$  is the maximum value of any component of $S_{n'n}$.
Factors of $i$ have been absorbed into $\oo x^m$, $\oo p^m$,
and $\oo L_{m'm}$ so that they are anti-Hermitian with respect to the spinor form $\beta$
of $\3S^6$.

This quantizes not only space-time $\{x^{\mu}\}$ 
but the orbital space $\{x^{\mu}, p_{\mu}, L_{\mu'\mu}, i\}$
as well.
Phase space and orbital space are already quantum spaces in the standard quantum theory,
in that momentum and position do not commute.
The singular canonical commutation relations of Heisenberg among 
the operators $x, p, L, i$ of ordinary relativistic mechanics
are all limits of regular angular momentum commutation relations
of the usual form $[L,L]\sim L$ among components of the higher-dimensional Yang angular momentum $L_{y'y}$.

\section{\lbl{S:DISCUSSION} Discussion}

These assignments of physical meaning to some quantum sets
are the opposite of experimental prediction.
They do not extract  experimental information from the theory
but put it in,
an indispensable first phase of theory construction.

Since set theory is a universal language, 
it is hardly surprising that its quantum version can be used 
to express parts of the Standard Model,
but the expression turns out simpler 
than we had any right to expect,
unless the theory has some correctness. 
If it extends at least approximately to the rest of the Standard Model, it removes some of the mystery of the Standard Model at the same time that it regularizes it.
It answers Rabi's question about the muon, ``Who ordered this?".
What ordered the muon
is Nature's modular architecture.

A finite quantum set theory of the orbital variables of the fermion is already known.
The synthesis of the orbital and internal algebras is under study.

\section{\lbl{S:ACKNOWLEDGMENT} Acknowledgment}

I owe S. Alexander, G. D'Ariano, G. F. R. Ellis, S. R. Finkelstein,  
Tenzin Gyatso, H. Saller, and F. Tony Smith
for discussions and information, and 
FQXi, the Templeton Foundation, and Dartmouth College
for supporting some of the presentations of this work.

\end{document}